\documentclass[preprint,5p,twocolumn]{elsarticle}

\usepackage{amssymb}
\usepackage{graphicx}
\usepackage{amssymb}
\usepackage{amsfonts}
\usepackage{dcolumn}% Align table columns on decimal point
\usepackage{bm}% bold math
\usepackage{lipsum} % to fill up some random text. 

\usepackage{xcolor} % different colored text for comments
\usepackage{caption}
\usepackage{array}
\usepackage[
  separate-uncertainty = true,
  multi-part-units = repeat
]{siunitx}
\usepackage{tabularx}

\usepackage{textgreek}
\usepackage{mathtools}
\usepackage{amsmath}

\journal{Astroparticle Physics}

\begin{document}

\begin{frontmatter}



\title{Study on NaI(Tl) crystal at \num{-35}$^{\circ}$C for dark matter detection}


\author[a,b]{S.~H.~Lee}

\author[c]{G.~S.~Kim}
\author[c]{H.~J.~Kim}
\author[b]{K.~W.~Kim}
\ead{kwkim@ibs.re.kr}
%\address[label1]{Center for Underground Physics, Institute for Basic Science (IBS), Daejeon 34126, South Korea}

\author[c]{J.~Y.~Lee}
\author[b,a]{H.~S.~Lee}
\ead{hyunsulee@ibs.re.kr}
%\affiliation{Center for Underground Physics, Institute for Basic Science (IBS), Daejeon 34126, South Korea}
%\affiliation{University of Science and Technology (UST), Daejeon 34113, South Korea}

\address[a]{University of Science and Technology~(UST), Daejeon 34113, South Korea}
\address[b]{Center for Underground Physics, Institute for Basic Science~(IBS), Daejeon 34126, South Korea}
\address[c]{Department of Physics, Kyungpook National University, Daegu 41566, South Korea}

\begin{abstract}
		We present the responses of a NaI(Tl) crystal in terms of the light yield and pulse shape characteristics of nuclear recoil events at two different temperatures: 22$^{\circ}$C (room temperature) and \num{-35}$^{\circ}$C (low temperature).  
The light yield is measured using 59.54\,keV $\gamma$-rays using a $^{241}$Am source relative to the mean charge of single photoelectrons. 
At the low temperature, we measure a 4.7 $\pm$ 1.3\% increase in the light yield compared to that at room temperature.
A significantly increased decay time is also observed at the low temperature. 
The responses to nuclear recoil events are measured using neutrons from a $^{252}$Cf source and compared to those to electron recoil events. 
The measured pulse shape discrimination (PSD) power of the NaI(Tl) crystal at the low temperature is found to be improved in the entire energy range studied because of the increased light yield and the different scintillation characteristics. 
We also find an approximately 9\% increased quenching factor of $\alpha$-induced events, which is the light yield ratio of $\alpha$ recoil to electron recoil, at the low temperature.  
This supports the possibility of an increased quenching factor of the nuclear recoil events that are known to have similar processes of dark matter interaction. 
The increased light yield and the improved PSD power of nuclear recoil events enhance the sensitivity for dark matter detection via dark matter--nuclei interactions. 
\end{abstract}




\end{frontmatter}


\section{Introduction}
\label{sec:intro}
Numerous astrophysical observations provide evidence that the dominant matter in the universe is dark matter~\cite{Komatsu:2010fb,Ade:2013zuv}. A weakly interacting massive particle (WIMP) is one of the stringent candidates of particle dark matter~\cite{lee77,jungman96}. 
In the last few decades, there have been many efforts on the direct detection of WIMP dark matter without success~\cite{Aprile:2018dbl,PhysRevLett.118.021303,Agnese:2017njq,Tanabashi:2018oca}. 
However, one noticeable exception is the annual modulation signal observed in the DAMA/LIBRA experiment using an array of NaI(Tl) detectors~\cite{Bernabei:2018yyw}, which could be interpreted as a WIMP--nucleon interaction~\cite{Baum:2018ekm,Ko:2019enb}. 
Several experimental groups have been designing new experiments with the aim of reproducing or disproving the results of the DAMA experiment using NaI(Tl) detectors~\cite{Amare:2018ndh,sabre,Angloher:2017sft,Adhikari:2019off}.
In the COSINE-100~\cite{Adhikari:2018ljm,Adhikari:2019off,COSINE-100:2021xqn} and ANAIS-112~\cite{Amare:2019jul,Amare:2021yyu} experiments, null signals were found; however, these results are yet to reach the 3$\sigma$ level conclusion in model-independent annual modulation studies. 

In contrast, experiments with liquid xenon have almost reached neutrino levels in the WIMP mass and spin-independent WIMP-proton cross-section parameter spaces~\cite{Aprile:2018dbl,PhysRevLett.118.021303}, which are approximately seven orders of magnitude smaller than the allowed regions from the DAMA results interpreted using the same model~\cite{Tanabashi:2018oca,Schumann:2019eaa}. 
However, there is still an uncovered region using liquid xenon in which the WIMP mass is low, below 5\,GeV, and the WIMP--proton interaction is spin-dependent ~\cite{Tanabashi:2018oca}. 
In addition to reproducing the DAMA results, NaI(Tl) crystals are particularly interesting owing to their high light yield observed up to 22\,photoelectrons per keV~\cite{Choi:2020qcj} and the sensitivity of the odd number of protons in both sodium and iodine to spin-dependent WIMP--proton interactions. 
Because ultra-pure NaI(Tl) crystals are produced well by many groups for reproducing DAMA/LIBRA experiments ~\cite{Suerfu:2019snq,Park:2020fsq,Fushimi:2021mez}, future dark matter search experiments using NaI(Tl) crystals can explore the uncovered parameter space of WIMP dark matter~\cite{Adhikari:2021kwv}. 
Increasing the light yield and improving the pulse shape discrimination (PSD) power for nuclear recoil events are essential for increasing sensitivity of such experiments. 

Temperature dependence of the scintillation responses of NaI(Tl) crystals was reported in ~\cite{Ianakiev:2006bi,Sailer:2012ua,Sibczy_ski_2012,Lee:2021jfx}. The increase in the light yield from room temperature to a low temperature until \num{-35}$^{\circ}$C~\cite{Sailer:2012ua,Lee:2021jfx} is the main motivation to study these specific temperatures.
The literature also reports an increased scintillation decay time at a low temperature~\cite{Ianakiev:2006bi,Sailer:2012ua,Lee:2021jfx}. 
The PSD between nuclear recoil events, which can be produced by WIMP--nuclei interactions, and electron recoil events, which are caused by unwanted $\gamma$ and $\beta$ backgrounds, depend on the scintillation decay times for different recoils~\cite{Kim:2018kbs}. However, the responses of a NaI(Tl) crystal to nuclear recoil events at low temperatures have not been measured. 
In this study, we compare the detection performance of a NaI(Tl) crystal at \num{-35}$^{\circ}$C to that at room temperature 22$^{\circ}$C and examine its feasibility as a future dark matter search detector. 

\section{Experimental setup}
\label{sec:setup}
\begin{figure}[tbp]
\centering
\begin{tabular}{cc}
\includegraphics[width=0.22\textwidth]{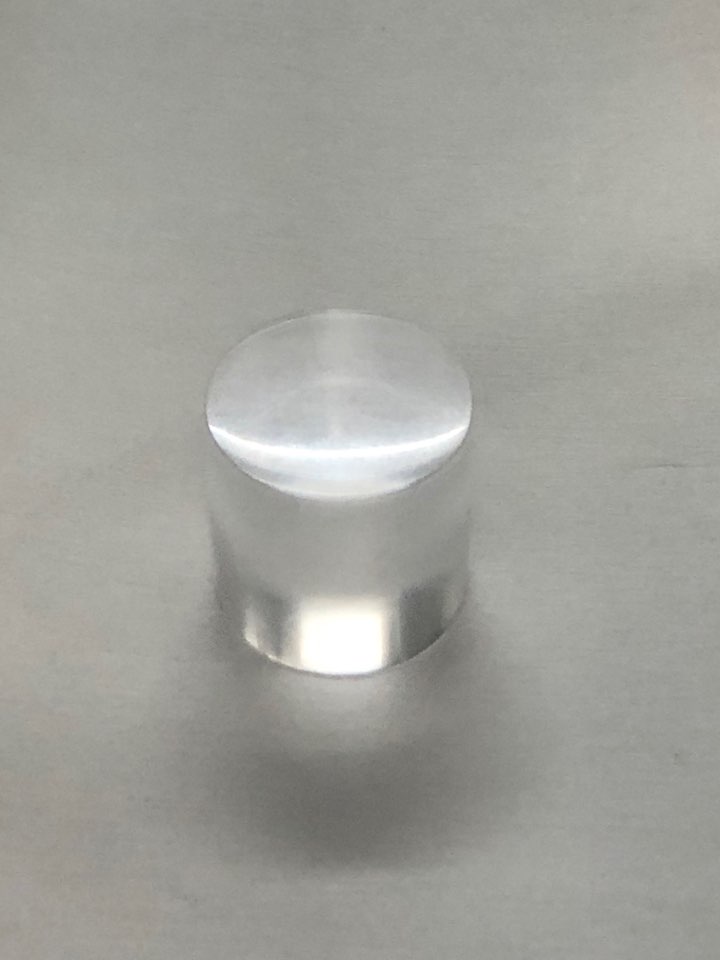} &
\includegraphics[width=0.22\textwidth]{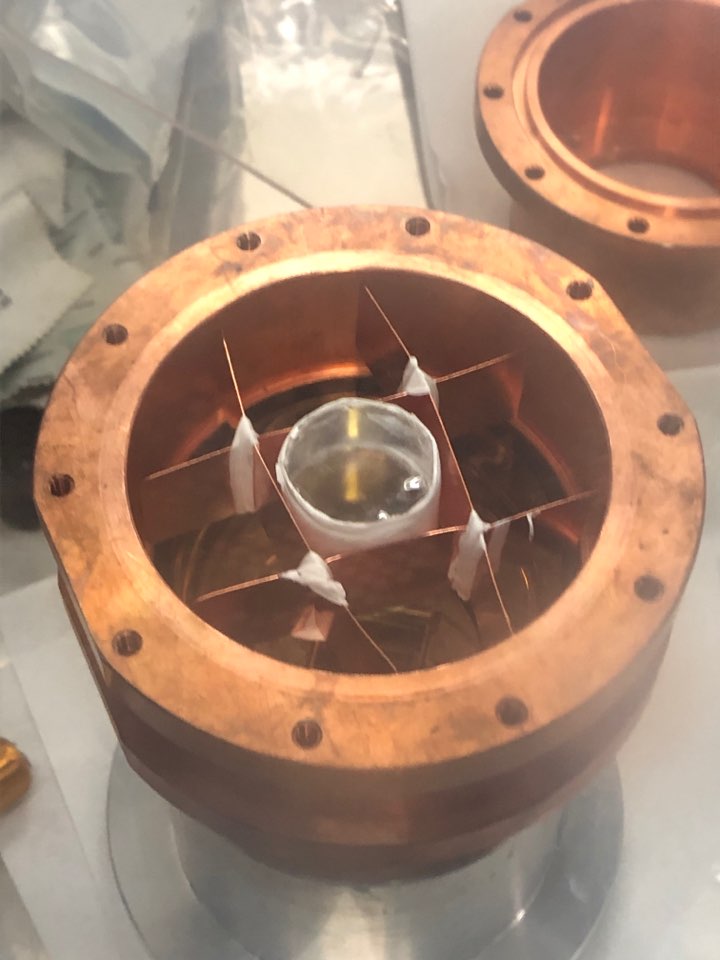} \\
(a) & (b) \\
\end{tabular}
\caption{ 
(a) NaI(Tl) crystal with dimensions of 2.48\,cm diameter and 3.77\,cm height is polished in glove box. 
(b) Crystal is installed inside copper encasing for assembling with PMTs. 
}
\label{fig_setup}
\end{figure}

A cylindrical-shaped NaI(Tl) crystal with dimensions of 2.48\,cm diameter and 3.77\,cm height from Alpha Spectra Inc. is used in the experiments, as shown in Fig.~\ref{fig_setup} (a).
The crystal was cut from the same ingots used for the COSINE-100 experiment~\cite{Adhikari:2017esn}, called C6 and C7.
As one can see in Fig.~\ref{fig_setup} (b), the crystal is centered in a copper structure and encased in a copper housing. 
Two 3-inch photomultiplier tubes (PMTs) (model number R12669SEL, Hamamatsu Photonics) selected for their high quantum efficiency are directly attached to two ends of the crystal.
An O-ring between the barrel of the PMT and the copper encasing seals the crystal from humidity in air. 
Only a single optical pad between the PMT window and the NaI(Tl) end surface maximizes the light collection efficiency, as described in Ref.~\cite{Choi:2020qcj}. 
The crystal was assembled as a detector inside a low-humidity N$_2$-gas flushed glove box in which the humidity level was maintained to be less than a few tens of ppm (H$_2$O) using a molecular sieve trap~\cite{Park:2020fsq}. 
All assembly parts were cleaned using dilute Citranox liquid by sonication, and baked in an oven before being moved into the glove box. 

A commercial freezer was used to achieve a detector temperature of \num{-35}$^{\circ}$C. Lead bricks of 5\,cm thickness surrounded the detector in the freezer to prevent the external background. 
The room temperature (22$^{\circ}$C) measurement was also conducted in the same setup but without the operation of the freezer.
Signals from the PMTs attached at each end of the NaI(Tl) crystal were amplified by home-made preamplifiers and digitized by a 500\,MHz, 12-bit flash analog-to-digital converter. 
A trigger was generated when a signal corresponding to one or more photoelectrons occurred in each PMT within a 200\,ns time window. 
An 8\,$\mu$s-long waveform starting 2.4\,$\mu$s before the time of the trigger position was stored for the triggered events at room temperature. 
However, because of the increased scintillation decay time at the low temperature, we acquired a 16\,$\mu$s-long waveform at \num{-35}$^{\circ}$C. 

A response of the crystal to an electron recoil event is calibrated by 59.54\,keV $\gamma$-rays emitted from an $^{241}$Am source. 
A nuclear recoil response is measured using neutrons emitted from a $^{252}$Cf source. 
For the measurements at the different temperatures, we placed the source at the same location to avoid any side effects due to a position dependence of the NaI(Tl) crystal. 
Intrinsic contamination of $^{210}$Pb resulting in an $\alpha$-decay of $^{210}$Po inside the crystal was used to evaluate the response of $\alpha$-particles. 

\section{Data analysis and Results}
\subsection{Decay time measurement}

\begin{figure}[tbp]
\centering
\includegraphics[width=0.9\columnwidth]{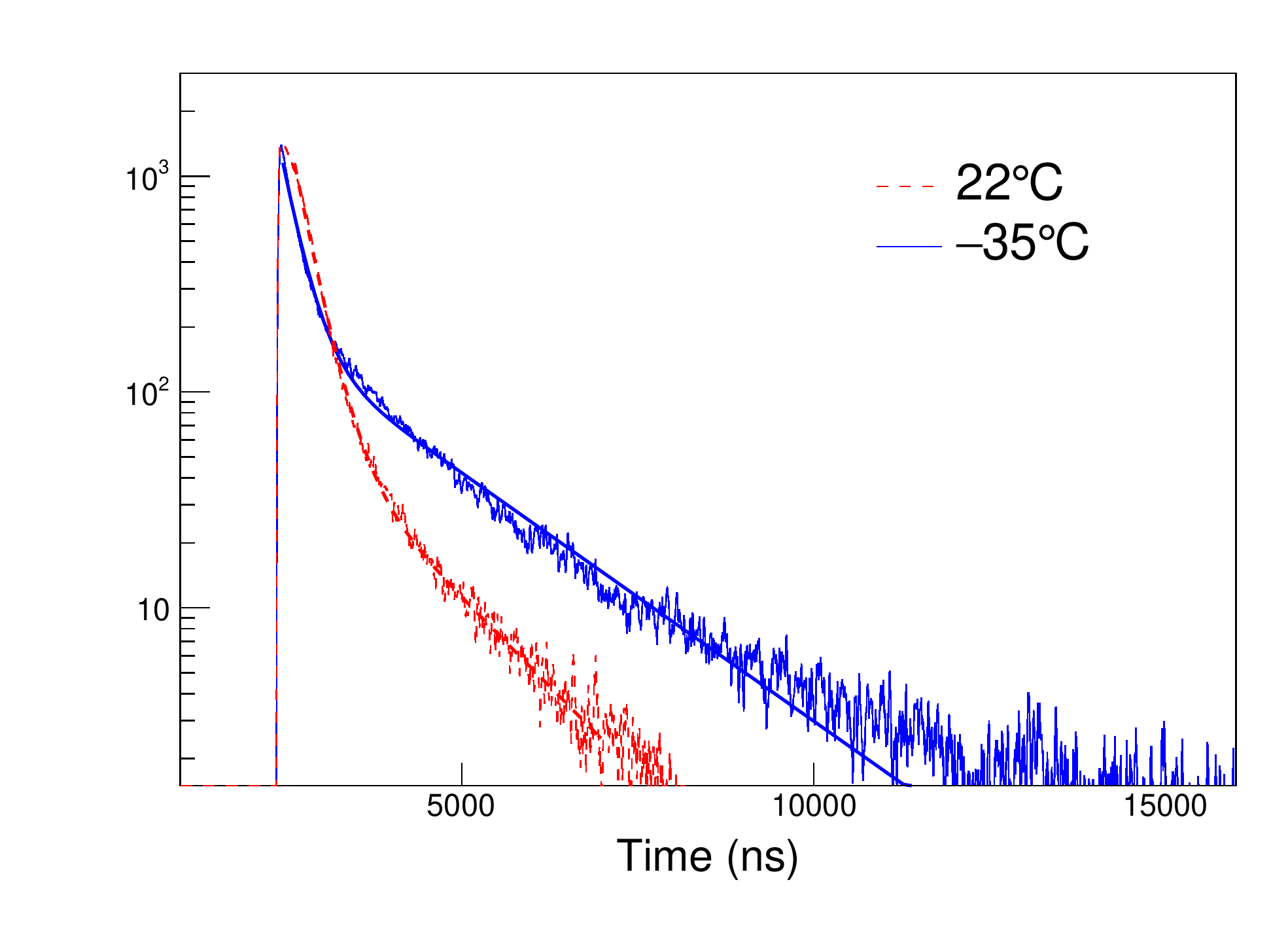}
\caption{
Accumulated waveforms of 59.54\,keV-$\gamma$ events at two different temperatures are fitted by two exponential functions. 
Red dashed line and blue solid line represent waveforms at 22$^{\circ}$C and \num{-35}$^{\circ}$C, respectively.
}
\label{fig_wav}
\end{figure}

We study the scintillation decay time of the NaI(Tl) crystal from its time response to the signals induced by the radioactive $\gamma$-ray source. 
The measured waveforms from 59.54\,keV $\gamma$-rays for the two different temperatures are integrated, and their comparison is shown in Fig.~\ref{fig_wav}. 
The data are fitted with two exponentials, and the results are summarized in Table~\ref{table-1}, where $\tau_1$ and $\tau_2$ are the decay constants of the two exponential fits for fast and slow components, respectively. 
We find that the fast decay components have a similar decay constant of approximately 250\,ns at both 22$^{\circ}$C and \num{-35}$^{\circ}$C.
However, a significantly longer slow decay constant at \num{-35}$^{\circ}$C of approximately 1.9\,$\mu$s than that at room temperature, 1.3\,$\mu$s, is observed. 
The relative amount of the fast decay components is also significantly decreased at the low temperature, as summarized in Table~\ref{table-1}. 
An increased decay time at low temperatures was already reported in ~\cite{Ianakiev:2006bi,Sailer:2012ua,Lee:2021jfx}.

\begin{table}[h!]
  \begin{center}
			\caption{Fitted parameters of accumulated waveforms in Fig.~\ref{fig_wav}. Fast decay constant $\tau_1$, slow decay constant $\tau_2$, and relative amount of fast component (\% of $\tau_1$) at two temperatures are summarized. 
			}
    \label{table-1}
      \begin{tabular}{c|c|c|c}
					\hline
      \textbf{Temp. ($^{\circ}$C)}  & \textbf{${\tau}_{1}$ (ns)} & \textbf{${\tau}_{2}$ (ns)} & \textbf{\% of ${\tau}_{1}$}\\
      \hline
      {22} & \SI{258 \pm 2}{} &  \SI{1320 \pm 19}{} & \SI{78 \pm 5}{} \\
      {\num{-35}} & \SI{245 \pm 2}{} &  \SI{1888 \pm 9}{} & \SI{51 \pm 3}{} \\
					\hline
    \end{tabular}
  \end{center}
\end{table}

\subsection{Light yields}

\begin{figure}[tbp]
\centering
\begin{tabular}{c}
\includegraphics[width=0.45\textwidth]{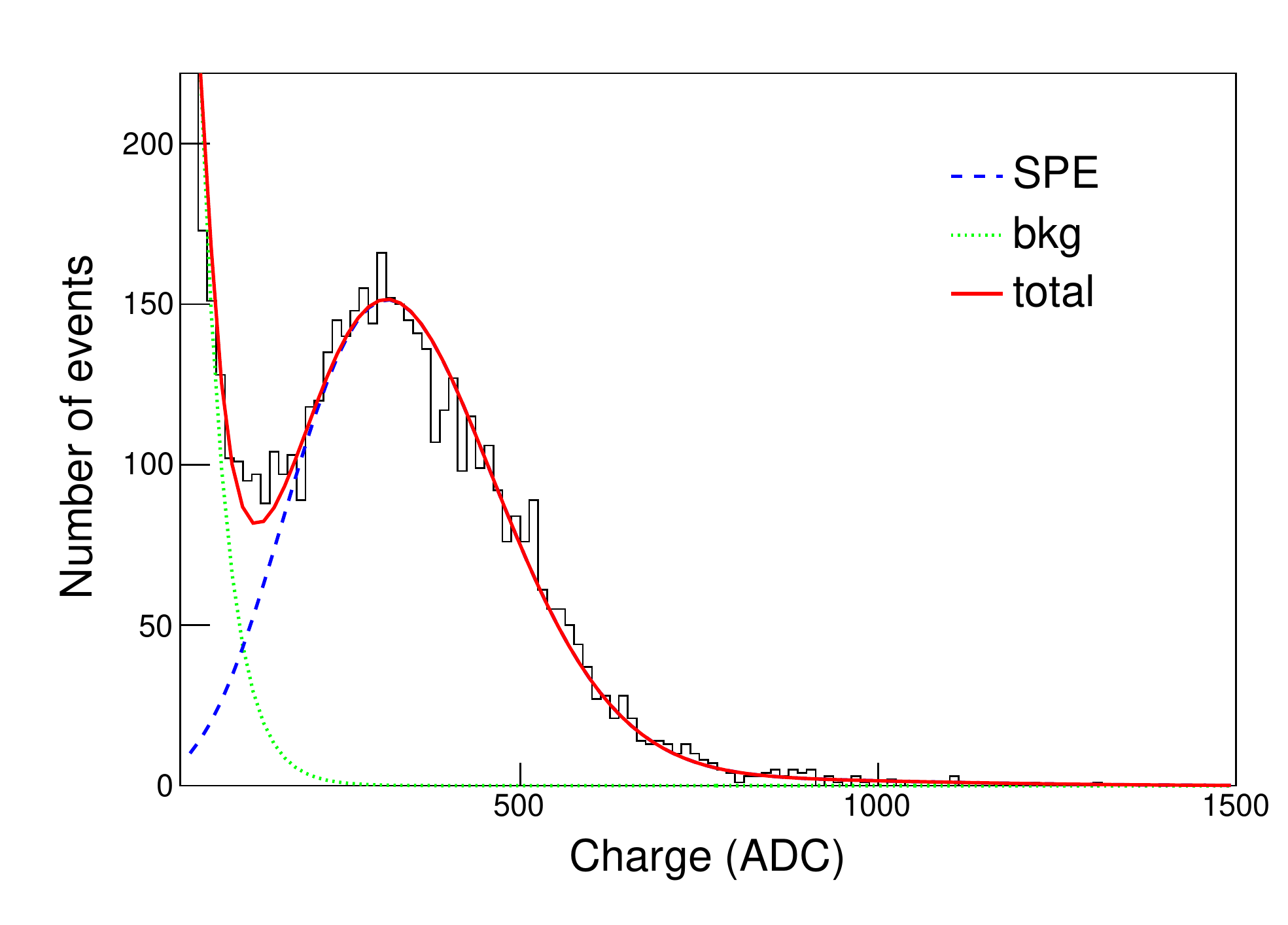} \\
(a) \\
\includegraphics[width=0.45\textwidth]{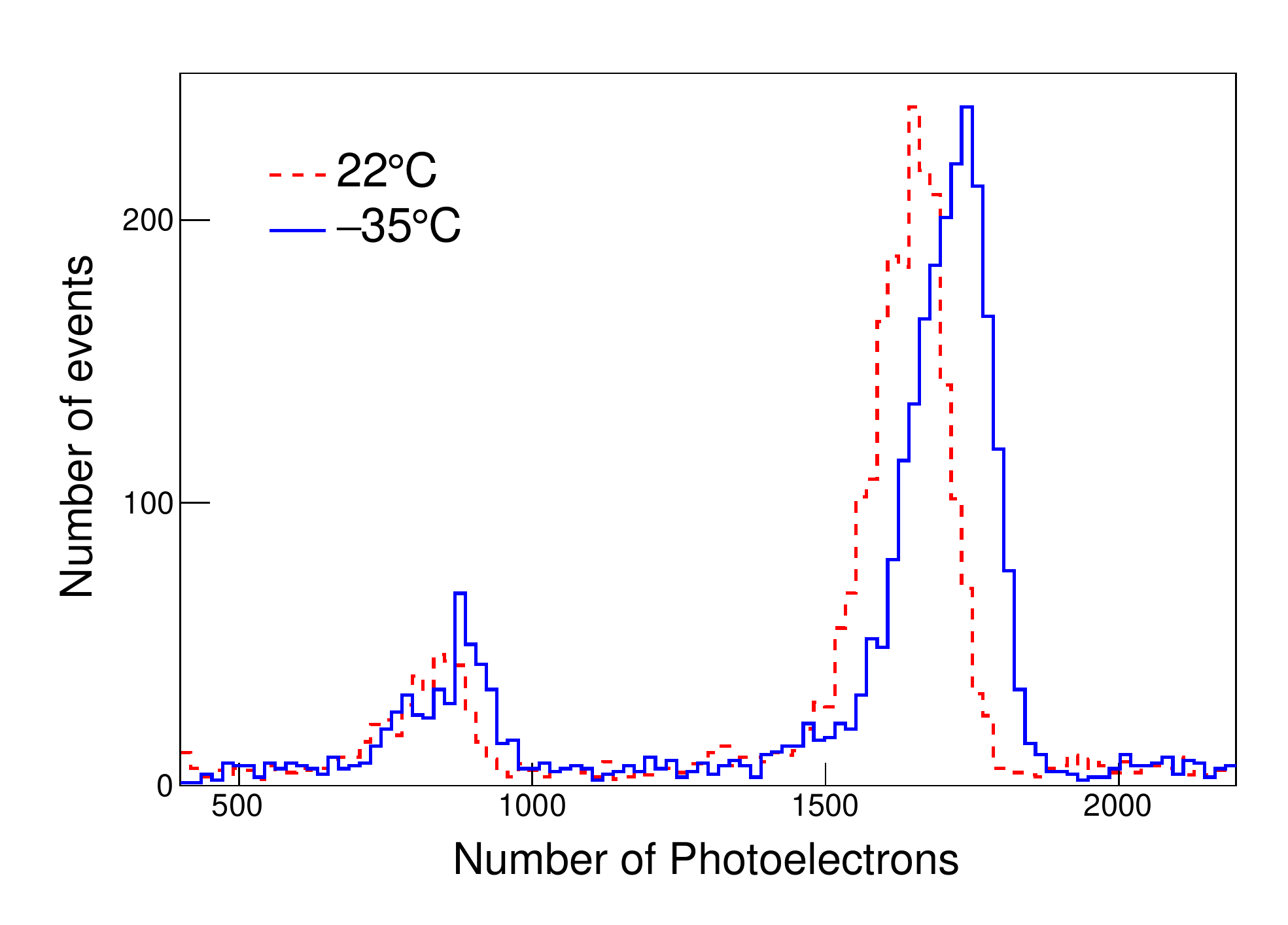} \\
(b) \\
\end{tabular}
\caption{ \label{fig_spe_npe}
(a) Charge distribution of SPEs measured at 22$^{\circ}$C is modeled with exponential background (green dotted line) and Poisson signal (blue dashed line).
(b) Number of photoelectrons for 59.54\,keV $\gamma$-rays from $^{241}$Am source is shown for two different temperature measurements at 22$^{\circ}$C (red dashed line) and \num{-35}$^{\circ}$C (blue solid line).
}
\end{figure}

We identify single photoelectrons (SPEs) using isolated clusters at the decay tail of a signal to suppress multiple photoelectron clusters~\cite{Lee:2005qr}. 
We adjust the range of the tail window to occupy similar number of SPEs considering the different decay times at the different temperatures.  
Figure~\ref{fig_spe_npe} (a) shows the charge distribution of the identified SPEs at room temperature as an example. 
The SPE distribution is fitted with a function (red solid line) that is the sum of a Poisson signal (blue dashed line) and an exponential background (green dotted line).
The summed charge distribution for the 59.54\,keV $\gamma$-rays is normalized with the mean of the SPE charge, as shown in Fig.~\ref{fig_spe_npe} (b). 
Here, we use different integration time ranges to contain 95\% of the total charge by modeling the waveform in Fig.~\ref{fig_wav} and Table~\ref{table-1}. 
The integration ranges are determined as 0.97\,$\mu$s and 4.8\,$\mu$s from the trigger positions at 22$^{\circ}$C and \num{-35}$^{\circ}$C, respectively.
It can be seen from Fig.~\ref{fig_spe_npe} (b), the \num{-35}$^{\circ}$C measurement achieves an approximately 5\% increased light yield, which is consistent with previous reports~\cite{Ianakiev:2006bi,Sailer:2012ua}. 

The measured light yields per keV are calculated, and the results are summarized in Table~\ref{table-2}.
Owing to the improved light collection by the direct attachment of the PMTs, as described in Ref.~\cite{Choi:2020qcj}, and the small crystal size, 
we obtain an extremely high light yield per keV of 27.6 $\pm$ 0.3\,PEs/keV at room temperature. 
At \num{-35}$^{\circ}$C, it is increased to 28.9 $\pm$ 0.2\,PEs/keV. 
This is the highest measured light yield of a NaI(Tl) crystal reported until now.

\begin{table}[h!]
  \begin{center}
			\caption{Measured light yields at two different temperatures using 59.54\,keV $\gamma$-rays and SPE distribution in Fig.~\ref{fig_spe_npe}.}
     \label{table-2}
      \begin{tabular}{c|c|c}
      \hline
      \textbf{Temp. ($^{\circ}$C)} & \textbf{LY (PEs/keV)} & \textbf{$\sigma$/mean (\%)} \\ %& \textbf{NPE in $^{241}$Am peak} 
      \hline
      {22} & \SI{27.6 \pm 0.3}{} & \SI{3.8 \pm 0.1}{} \\ %& \SI{1595.2 \pm 3.1}{} 
      {\num{-35}}  & \SI{28.9 \pm 0.2}{} & \SI{3.7 \pm 0.1}{} \\ %&\SI{1677.1 \pm 2.1}{}
      \hline
    \end{tabular}
  \end{center}
\end{table}

%\subsection{Pulse shape discrimination between electron recoil and nuclear recoil events}
\subsection{PSD between electron and nuclear recoil events}
The scintillation characteristics of a NaI(Tl) crystal are different for nuclear and electron recoils~\cite{Gerbier:1998dm,Lee:2015iaa}, which has allowed searching for WIMP--nuclei interaction events based on PSD analysis~\cite{Kim:2018wcl}. 
Improving the PSD between nuclear and electron recoils is directly related to the improvement in the sensitivity of WIMP dark matter search~\cite{Lee:2015iaa}. 
Therefore, the increase in both the decay time and light yield in the low-temperature measurements of the PSD power is interesting. 

\begin{figure}[tbp]
\centering
\begin{tabular}{c}
\includegraphics[width=0.45\textwidth]{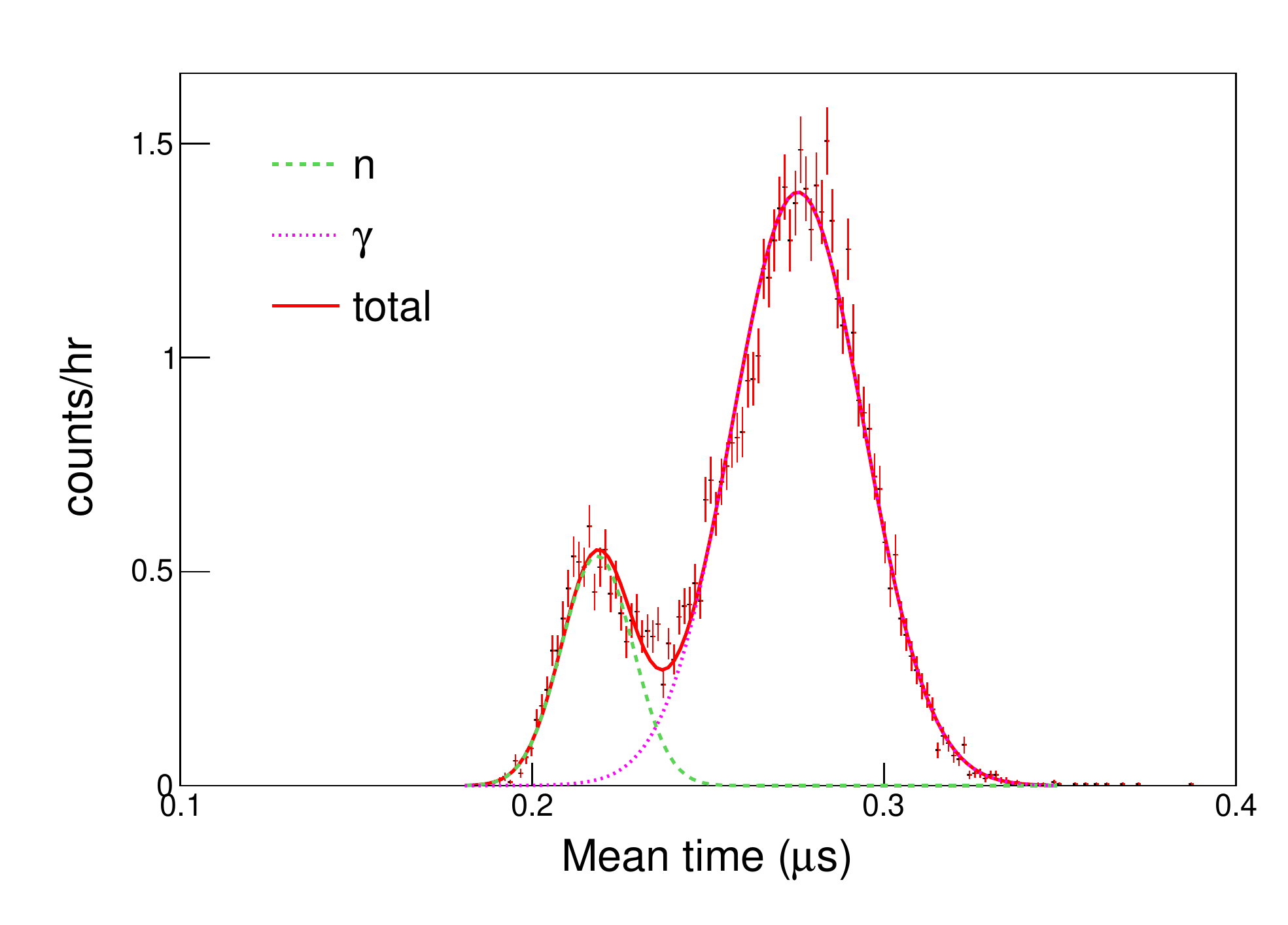} \\
(a) \\
\includegraphics[width=0.45\textwidth]{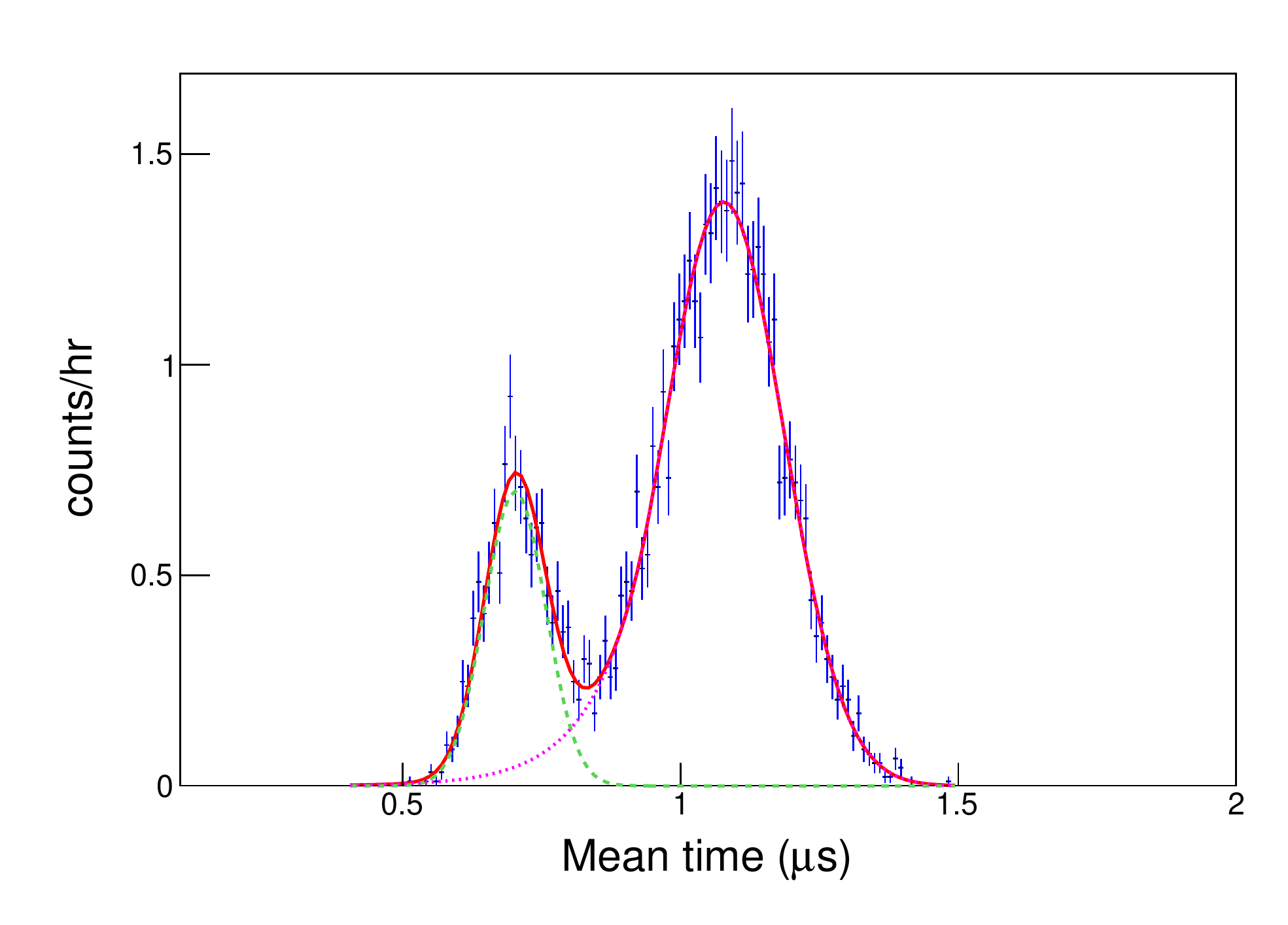} \\
(b) \\
\end{tabular}
\caption{ \label{fig_mt}
Mean time distributions of neutron calibration data using $^{252}$Cf source at energies from 20 to 25 keV electron equivalent for 22$^{\circ}$C (a) and \num{-35}$^{\circ}$C (b) are presented. Magenta dotted line and green dashed line correspond to electron and nuclear recoil events.
}
\end{figure}

For nuclear recoil events, we irradiated neutrons from the $^{252}$Cf neutron source to the NaI(Tl) crystal. 
The mean decay time defined as 
\begin{equation}
\label{eq_mt}
\text{Mean~Time} \equiv \Big( \frac{\sum_{i=1}^n h_{i}t_{i}}{\sum_{i=1}^n h_{i}} -t_{0}\Big)
\end{equation}
is used to quantify the PSD power. 
Here, $h_{i}$ and $t_{i}$ are the height and time of the $i$th bin, respectively, and $t_{0}$ is the time of the rising edge above the threshold. 
We obtain the mean time distributions for each 5\,keV bin from 10\,keV to 60\,keV at the two different temperatures. 
The mean time distributions between 20 and 25\,keV are shown in Fig.~\ref{fig_mt} at 22$^{\circ}$C (a) and \num{-35}$^{\circ}$C (b) as examples.
The nuclear recoil events are modeled with the Gaussian function, whereas the electron recoil events by the Gaussian-convoluted crystal ball function. The tail in the mean time distribution may be caused by defects on the crystal surface, which cause inefficient scintillation for the low-energy $\gamma$ and $\beta$ events occurring on the crystal surface.  
The electron (dotted line) and nuclear (dashed line) recoils are distinguished by the model. 

\begin{figure}
     \centering
     \includegraphics[width=0.45\textwidth]{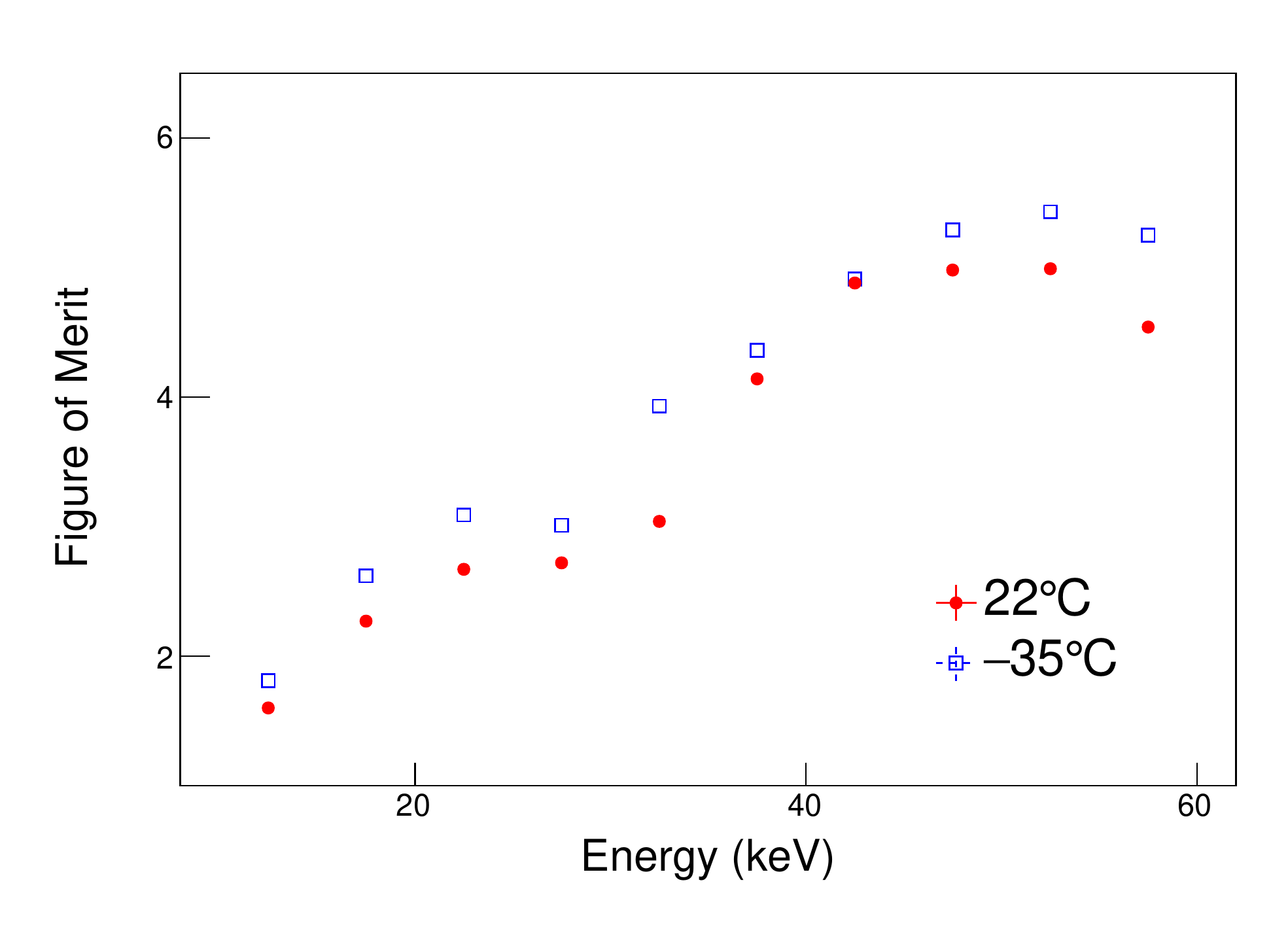}
%     \caption{Figure of merit between the nuclear recoil and the electron recoil events at 22$^{\circ}$C (red filled circles) and \num{-35}$^{\circ}$C (blue hollow squares) are presented.}
     \caption{FoMs between nuclear and electron recoil events at 22$^{\circ}$C (red filled circles) and \num{-35}$^{\circ}$C (blue hollow squares) are presented.}
     \label{fig-8}
\end{figure}

The PSD power can be quantified by a figure of merit (FoM) estimated using the mean and sigma values from the fitted models.
The FoM versus energy variations are presented in Fig.~\ref{fig-8} for 22$^{\circ}$C (red filled circles) and \num{-35}$^{\circ}$C (blue hollow squares) separately.
The \num{-35}$^{\circ}$C measurements provide improved PSD power than the room temperature measurements in the entire energy range.

\subsection{Alpha Quenching Measurement}
It is well known that scintillation light yields are different for electron and nuclear recoil events~\cite{Collar:2013gu,Xu:2015wha,Joo:2018hom}.
The quenching factor can be determined from the ratio of the measured electron equivalent energy to the true nuclear recoil energy, and is obtained as approximately 10$-$20\% for sodium and 4$-$6\% for iodine at nuclear recoil energies between 5 and 150\,keV~\cite{Joo:2018hom}. 
Considering the increased light yield for the same electron equivalent energy and the increased decay time at the low temperature, quenching factor measurement at the low temperature is particularly interesting. 
The measured energy spectra of the nuclear recoil events induced by the neutrons from the $^{252}$Cf source were studied; however, no noticeable differences were observed owing to the limited statistics and the environmental neutron background. 
In the future, the neutron calibration system using a deuterium--deuterium neutron generator and neutron tagging detectors described in Ref.~\cite{Joo:2018hom} can be used to measure the quenching factor of the nuclear recoil events at the low temperature. However, a setup of a freezer with a good alignment of neutron beams to the crystal and surrounding neutron tagging detectors requires additional efforts. 

\begin{figure}
    \centering
    \includegraphics[width=0.45\textwidth]{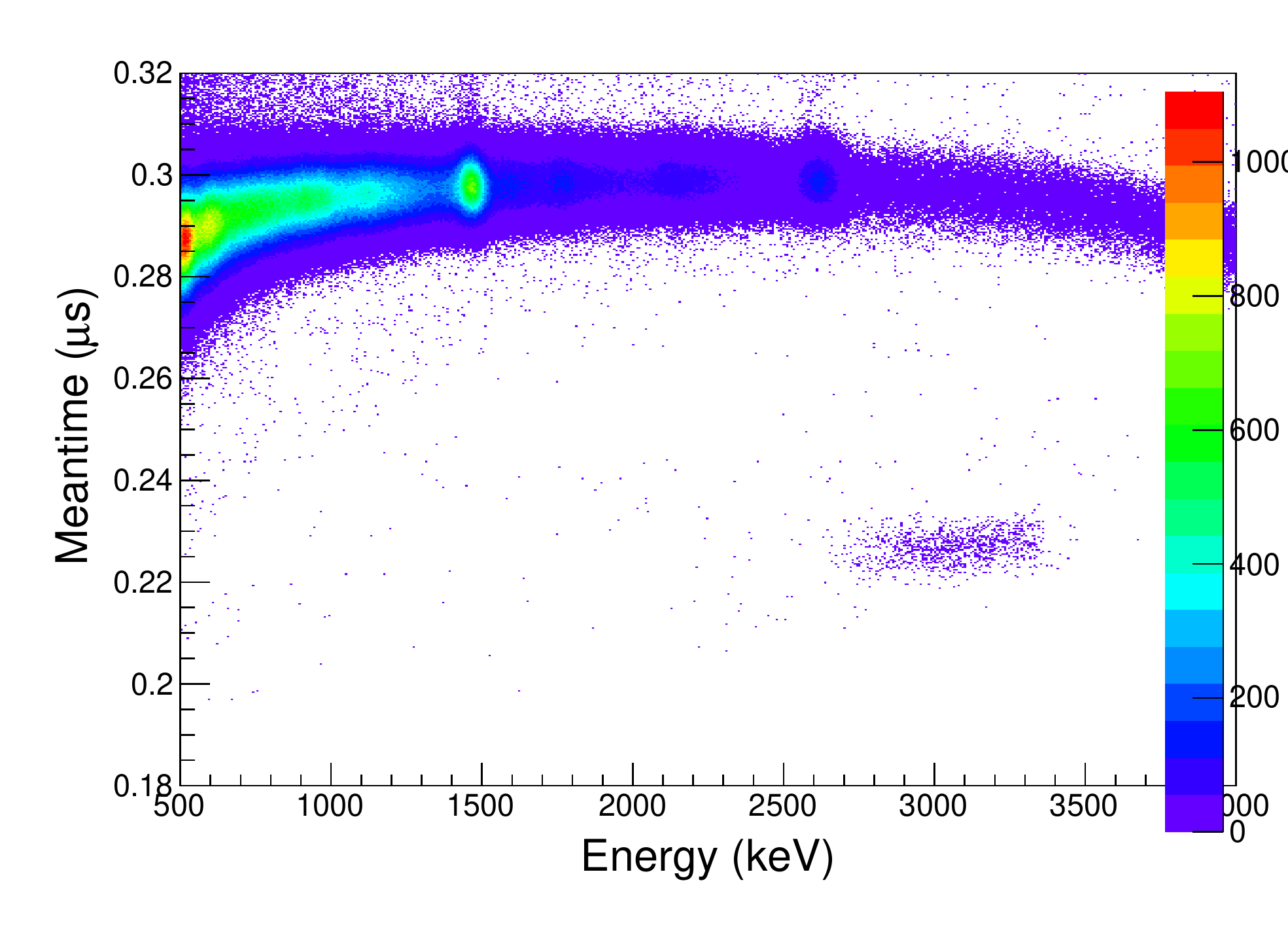}
    \caption{Energy versus mean time scattered plot using internal background of NaI(Tl) crystal at 22$^{\circ}$C is shown. Alpha-induced events are isolated well with short mean time. }
    \label{fig-10-emt}
\end{figure}

\begin{figure}[tbp]
\centering
\begin{tabular}{c}
\includegraphics[width=0.45\textwidth]{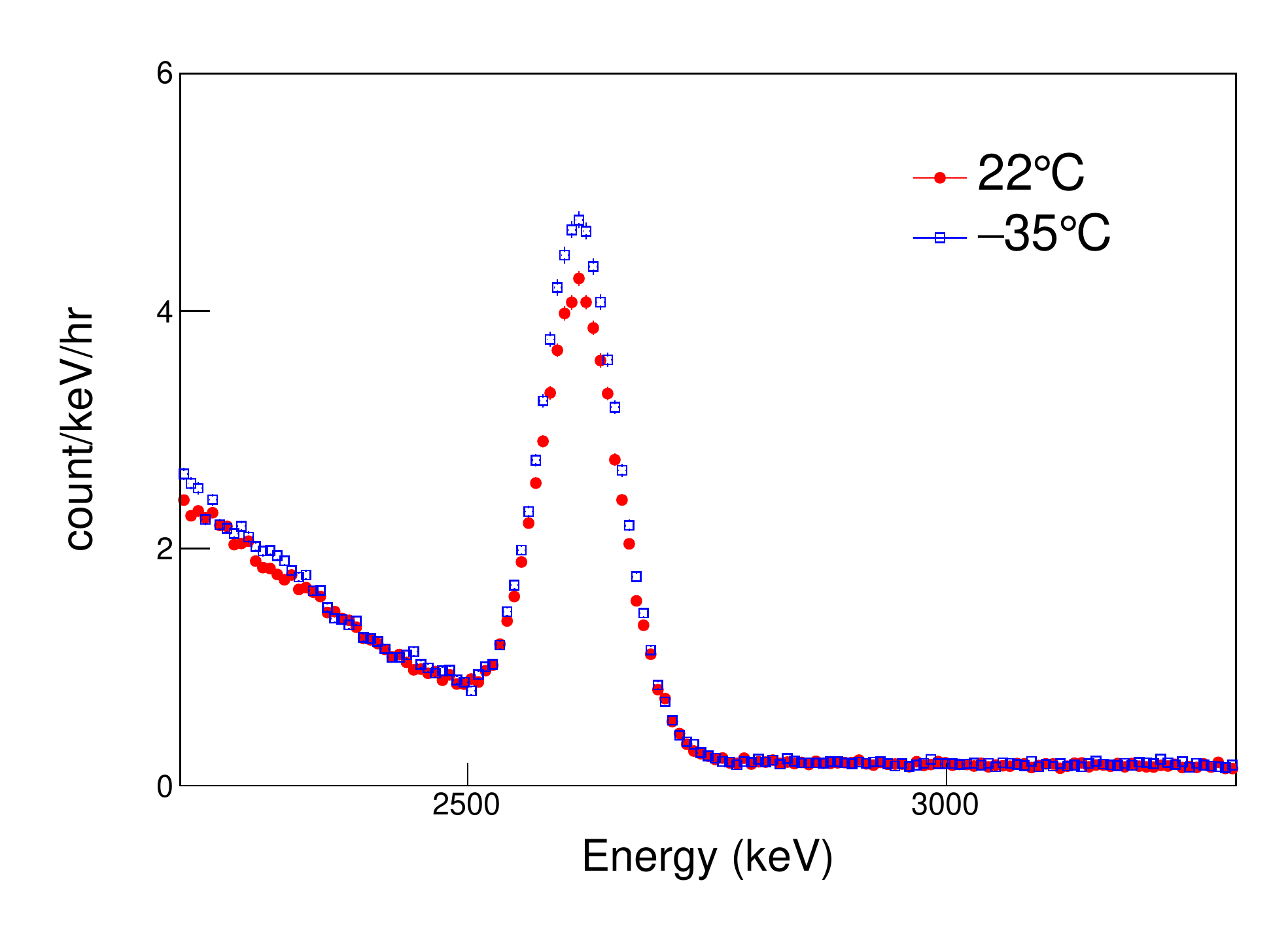} \\
(a) \\
\includegraphics[width=0.45\textwidth]{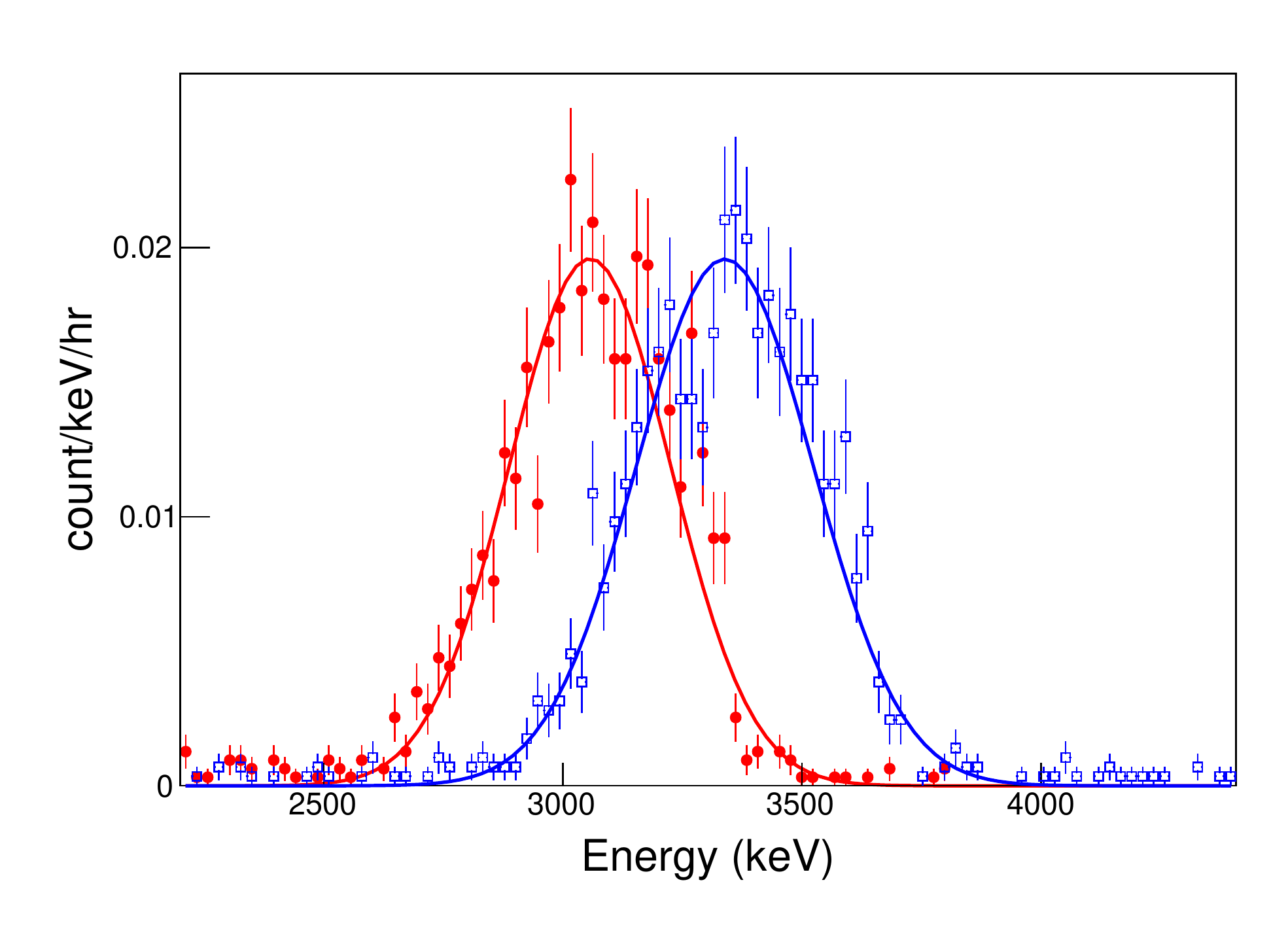} \\
(b) \\
\end{tabular}
\caption{ \label{fig_alpha_gamma}
Energy spectra of $\gamma$ and $\beta$-induced events (a) and $\alpha$-induced events (b) from internal background of NaI(Tl) crystal are presented for 22$^{\circ}$C (red filled circles) and \num{-35}$^{\circ}$C (blue hollow squares). 
}
\end{figure}

Owing to the difficulty in measuring the nuclear recoil quenching factor in the low-energy region, we studied the responses of $\alpha$-induced events originated from an internal contaminant of $^{210}$Pb, providing an $\alpha$ decay of $^{210}$Po with an $\alpha$ Q-value of 5.41\,MeV. 
$\alpha$-particles are easily identified by the mean time owing to their fast decay, similar to nuclear recoil events, as shown in Fig.~\ref{fig-10-emt}. 
Here, the electron-equivalent energies are calibrated with the 2.615\,MeV $\gamma$ line from $^{208}$Tl, as shown in Fig.~\ref{fig_alpha_gamma} (a). 
The energies of the $\alpha$-particles on the electron-equivalent scale are presented in Fig.~\ref{fig_alpha_gamma} (b). 
The electron-equivalent energies of the same $\alpha$ particles for the two different temperatures present noticeable deviations and are accounted as the temperature-dependent $\alpha$ quenching factor. 
We obtain a 9.23 $\pm$ 0.26\% increased $\alpha$ quenching factor at \num{-35}$^{\circ}$C compared to that at room temperature.  
Assuming similar responses for low-energy nuclear recoil events, the dark matter detection sensitivity of NaI(Tl) crystals operated at \num{-35}$^{\circ}$ temperature is expected to be significantly enhanced. 

\section{Conclusion}
We studied responses of a NaI(Tl) crystal at two different temperatures--- 22$^{\circ}$C and \num{-35}$^{\circ}$C---for dark matter detection. 
An increased light yield of 4.7 $\pm$ 1.3\% at \num{-35}$^{\circ}$C compared to that at 22$^{\circ}$C is observed for electron recoil events. 
In addition, an increased $\alpha$ quenching of 9.23 $\pm$ 0.26\% is measured, which provides a possibility of additional increased light yield for same-energy nuclear recoil events.
The PSD capabilities between nuclear and electron recoil events are evaluated by quantifying the FoM using the mean time distribution.
The improved PSD power at \num{-35}$^{\circ}$C will additionally enhance the dark matter detection sensitivity based on WIMP--nucleus interactions. 

\section*{Acknowledgments}
This work is supported by the Institute for Basic Science (IBS) under project code IBS-R016-A1.




\bibliographystyle{elsarticle-num}

%\begin{thebibliography}{00}


%\bibitem{}

%\end{thebibliography}

\end{document}